\def\ltsima{$\; \buildrel < \over \sim \;$}
\def\gtsima{$\; \buildrel > \over \sim \;$}
\def\simlt{\lower.5ex\hbox{\ltsima}}
\def\simgt{\lower.5ex\hbox{\gtsima}}

\documentstyle[11pt,paspconf,epsfig]{article}

\begin{document}

\title{X/$\gamma$-ray spectra of Seyferts and BeppoSAX observations}

\author{Giorgio Matt}
\affil{Dipartimento di Fisica, Universit\`a degli Studi "Roma Tre",
Roma, Italy}



\begin{abstract}
BeppoSAX results on Seyfert galaxies are reviewed and discussed.
\end{abstract}




\section{Introduction}

After more than 2 years of operation, it is time to
review the contribution that BeppoSAX is providing to our knowledge 
of Seyfert galaxies. As the best characteristic  of BeppoSAX is the wide
energy band (0.1--200 keV), the results will mainly concern broad
band spectra. Moreover, as shown in the last section, the
MECS instrument, which has the best spatial resolution
in the 2--10 keV band so far, is also giving valuable information 
on faint AGN and on the X--ray Background (XRB).

BeppoSAX (see Boella et al. 1997 for a comprehensive description of the 
mission) is composed by several instruments:
4 co--aligned Narrow Field Instruments (NFI); two Wide Field Cameras
pointing opposite each other and perpendicularly to the NFI; an
all sky $\gamma$-ray burst monitor. We are interested here in three
of the NFI: LECS, MECS and PDS. The MECS is presently composed of two units 
(after the failure of a third one in March 1997); the  working band is
1.5--10.5 keV, the
energy resolution is  $\sim$8\% and the angular resolution is
$\sim$0.7~arcmin (FWHM) at 6~keV. The characteristics of the LECS are 
similar to those of the MECS in the overlapping band, but its energy band
extends down to 0.1 keV. The PDS is a passively collimated detector
(about 1.5$\times$1.5 degrees f.o.v.) working in the 13--200 keV energy range.
A significant fraction of the BeppoSAX observing time (80\% the first year,
60\% the second year, 50\% afterwards) is devoted to a ``Core Program"
(reserved to collaborations led by Italian or Dutch PIs, with a fraction
also to MPE and ESA/SSD) which consists of major programs
aiming to observe classes of sources in a sistematic way.

In this paper I will assume that the unification model for Seyfert 
galaxies (Antonucci 1993) is valid, i.e. Seyfert 1 and 2 galaxies are intrinsically
identical (at least as far as the nuclear properties are concerned), 
and surrounded by a $\sim$pc-scale molecular
torus. If the line--of--sight does not intercept the torus, the nucleus
can be  directly observed and the source is classified as type 1; if
the line--of--sight is blocked by the torus, the source is classified as
type 2. I will come back to unification models in Sec.~2.

This review is necessarily
incomplete and inevitably biased towards my own interests. I will
discuss selected results on: bright, classical Seyfert 1 galaxies
(see Brandt, this volume, for results on Narrow Line Seyfert
1 galaxies); Seyfert 2 galaxies (expecially Compton--thick ones, 
and a sample of optically selected sources); the Seyfert 1 galaxy
NGC~4051, which
switched off in May 1998 and on again a month later;  and the HELLAS
program, i.e. a sample of hard X--ray selected sources, whose
identification program is in progress but which is already giving
valuable results on the X--ray background. 

\section{Seyfert 1 galaxies} 

Before describing BeppoSAX results on Seyfert 1 galaxies, let me briefly
recall the picture which has emerged after Ginga, ROSAT,
CGRO and ASCA observations (see e.g. Mushotzky, Done \& Pounds 1993;
Fabian 1996). 

The main component is a power law, cut--offing at high energies,
and very likely originating from Inverse Compton emission from 
relativistic electrons to UV/soft X--ray photons, the latter
possibly emitted by the accretion disc (see e.g. Svensson 1996 for a review). 

A significant fraction of the primary radiation is intercepted and 
reprocessed by optically thick matter, either the accretion disc or the torus,
or both. If the matter is neutral, the shape of the reflected component 
is determined basically by the competition between
photoelectric absorption (whose cross section depends on the energy, after each
photoabsorption edge,
as $\approx E^{-3}$) and Compton scattering (whose cross section is constant, 
at least up to a few tens of keV). For cosmic abundances,
the two cross sections are equal at about 10 keV.
This so--called Compton reflection component
has been studied in detail in several papers (see e.g. Lightman \& White 1988;
George \& Fabian 1991; Matt, Perola \& Piro 1991; Magdziarz \& Zdziarski
1995);  its spectrum is
a broad hump, peaking around 30 keV.
When added to the primary component, it hardens the total spectrum
above a few keV and steepens it above a few tens of keV.
Besides this Compton reflection continuum, the illumination of neutral matter
by the primary radiation results also in a strong iron 6.4 keV fluorescent
line,  emitted by iron atoms after removal of a K electron by
a X--ray photon. Kinematic and gravitational effects in the 
inner accretion disc modify 
the line profile (see e.g. Reynolds, this volume, and references therein),
as firstly observed by ASCA in MCG--6-30-15 (Tanaka et al. 1995) and later on
realized to be common in Seyfert 1s (Nandra et al. 1997).

At low energies (below $\sim$1 keV) a further component (``soft excess"), 
of rather unclear nature, may arise. Among the possible explanations, 
the tail of thermal emission from the accretion disc, and 
reflection from ionized matter (Ross \& Fabian 1993) are the most
popular. It is worth noticing that most
of the observations on soft excesses are based on relatively 
narrow band instruments (like ROSAT) or on non--simultaneous broad band
observations. 

All these components, which most likely originate close to the black hole,
(with the possible exception of the reflection from
the torus) may pass throughout ionized matter (the ``warm absorber").
The main signatures of this matter are absorption
edges of high ionization ions, mainly oxygen
(Halpern 1984; Nandra \& Pounds 1992; Fabian et al. 1994), which has
been unambiguosly observed by ASCA 
in a large fraction of Seyfert 1s (Reynolds 1997; George et al. 1997).
Resonant absorption lines may also be important (Matt 1994;
Krolik \& Kriss 1995; Nicastro, Fiore \& Matt 1998) and detectable by
future missions featuring high resolution spectrometers.

\subsection{A sample of bright Seyfert 1s}

Two major BeppoSAX Core programs are devoted to classical Seyfert 1s:
one (PI: G.C. Perola), aiming to study broad band spectra of bright sources;
the second (PI: L. Piro), looking for spectral variability. 
About a dozen sources have been observed on aggregate so far
(see Table~1). I am not going to
discuss in detail any of these sources, but I will rather try to 
outline the main results of Seyfert 1s as a class that are
emerging by these observations. Let me discuss separately the iron line,
the Compton reflection continuum, the high energy cut--off and the soft
excess.

\begin{table}
\caption{ BeppoSAX observations of bright Seyfert 1s. The second column
indicates whether a soft excess is required by the data (question marks
are for sources which are highly absorbed or, as for
 Fairall 9, the LECS was not in operation). The third
and four columns refer to the photon index of the primary power law
and the amount of 
reflection continuum, respectively (the latter calculated adopting a 
face--on slab); typical errors are of the orders of a few percent for
$\Gamma$, and $\simlt$50 percent for $R$. The fifth column reports the
$e$--folding energy, when an exponential cut--off is included 
(the two values for NGC~5506 are for the Jan 97 and Jan 98 
observations, respectively). For Mrk~766 PDS data have not been used
due to a confusion problems.
Cut--off errors are for $\Delta\chi^2$=4.61 (i.e. 90\%
confidence level for two parameters.}
\label{tbl-1}
\begin{center}
\begin{tabular}{lllll}
\tableline
\tableline
~& ~& ~ & ~ & ~\cr
Source & S.E. & $\Gamma$ & R & E$_c$~(keV) \cr
~& ~& ~ & ~ & ~\cr
\tableline
~& ~& ~ & ~ & ~\cr
NGC~4151$^1$  &      ?    &  1.2--1.5   & 0.1--0.5  &   70$\pm$15 \cr
NGC~5548$^2$  &      NO   &  1.55--1.65 &   0.7    &    200$\pm$140 \cr
NGC~5506$^3$  &      ?    &   2.05     &  1.1     &   $>$~400~($>$200) \cr
NGC~7469$^2$  &    Yes?   &  2.00 & 1.0      &  $>$~200 \cr
IC~4329A$^3$  &    NO      & 1.86       & 0.6      & 350$^{+470}_{-150}$ \cr
MCG-6-30-15$^4$     &  NO       & 2.05       & 1.3      &  170$^{+300}_{-70}$ \cr
NGC~4593$^5$  &    NO     & 1.87       & 1.1      &  $>$~150 \cr
Fairall~9$^3$ &    ?      & 2.01       & 0.9      &  $>$~260 \cr
NGC~3516      &    NO     & 1.77       & 1.3      &  $>$~100 \cr
Mrk~509       &    NO     & 1.94       & 1.7      & 105$^{+200}_{-50}$ \cr
Mrk~766       &  NO       & 2.15       & ?        &   ? \cr
~& ~& ~ & ~ & ~\cr
\tableline
\tableline
\end{tabular}
\end{center}
$^1$ Piro et al. 1998\par
$^2$ preliminary results: Piro, private communication\par
$^3$ Perola et al. 1998; Matt et al. 1998a\par
$^4$ Guainazzi et al. 1998a\par
$^5$ Guainazzi et al. 1998b\par
\end{table}

\subsubsection{Iron line.} 

Due to the poorer energy resolution of BeppoSAX with respect to ASCA,
no major advances about the iron line, especially when narrow, are to
be expected, even if the broad band may help better estimating the
underlying continuum. When the line is very broad, and the S/N good
enough,  like in the case of MCG--6-30-15, the MECS energy resolution is 
however sufficient to study the profile; this is demonstrated 
in Fig.~1, where the best fit model and residuals 
(with the line normalization set to zero for the sake of
illustration) are shown. The best fit parameters for the line are 
consistent with those of Tanaka et al. (1995). In less favourable cases
the line is usually still significantly broad, 
but the fit cannot discriminate between gaussian and relativistic disc
profiles; the relativistic line hypotesis is nevertheless to be preferred 
on plausibility arguments (Fabian et al. 1995).

\begin{figure}
\centerline{\epsfig{file=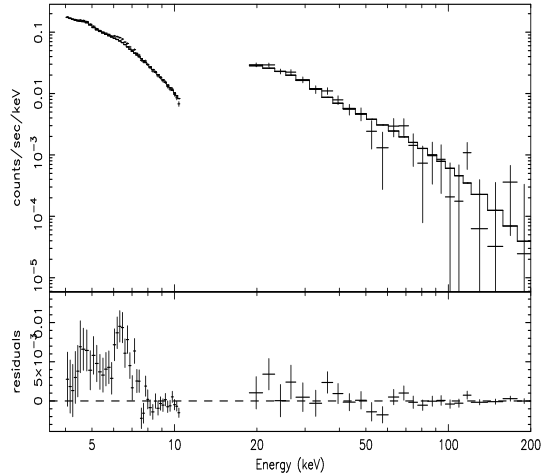,width=8cm,height=8cm,angle=-90}}
\caption{Spectrum and rediduals (with the line normalization set to zero for
illustration purposes) for MCG--6-30-15.}
\label{mcg6}
\end{figure}

\subsubsection{Compton reflection continuum.}

Where BeppoSAX is expected to improve previous observations is in the 
higher energy part of the spectrum, and in particular in the measurement
of the 
reflection component. The importance of the broad band in this respect is
best illustrated by Fig.~2 (left panel) where a model
with a reflection component included 
is fitted to the MECS data alone of NGC~5506 (first observation),
and then extrapolated to higher energies: the upper panel
shows still an excess of counts above 15 keV (due to the fact that
the reflection component is actually not required by the MECS data), 
which is well cured when the MECS and PDS data are fitted together
(lower panel). The right panel of Fig.~2 gives the
probability contours for the power law index, $\Gamma$,
and the relative normalization of the reflection component, 
R = $\Omega/2\pi$, where $\Omega$ is the solid angle
subtended by the reflecting matter at the illuminating source.
The two parameters are very strongly correlated,
and only the very broad energy band permits a good estimate of
$R$.

From Table 1, it is clear that the reflection component is ubiquitous 
in Seyfert 1s, so confirming the Ginga results (Nandra \& Pounds 1994). 
The values of $R$ in the Table have been 
estimated assuming a face--on slab, and therefore the actual values 
of $\Omega/2\pi$ may be somwaht higher. Assuming a typical error
of 30--50 percent, the measuread values are 
consistent with accretion disc reprocessing.

\begin{figure}
\begin{minipage}{60mm}
\epsfig{file=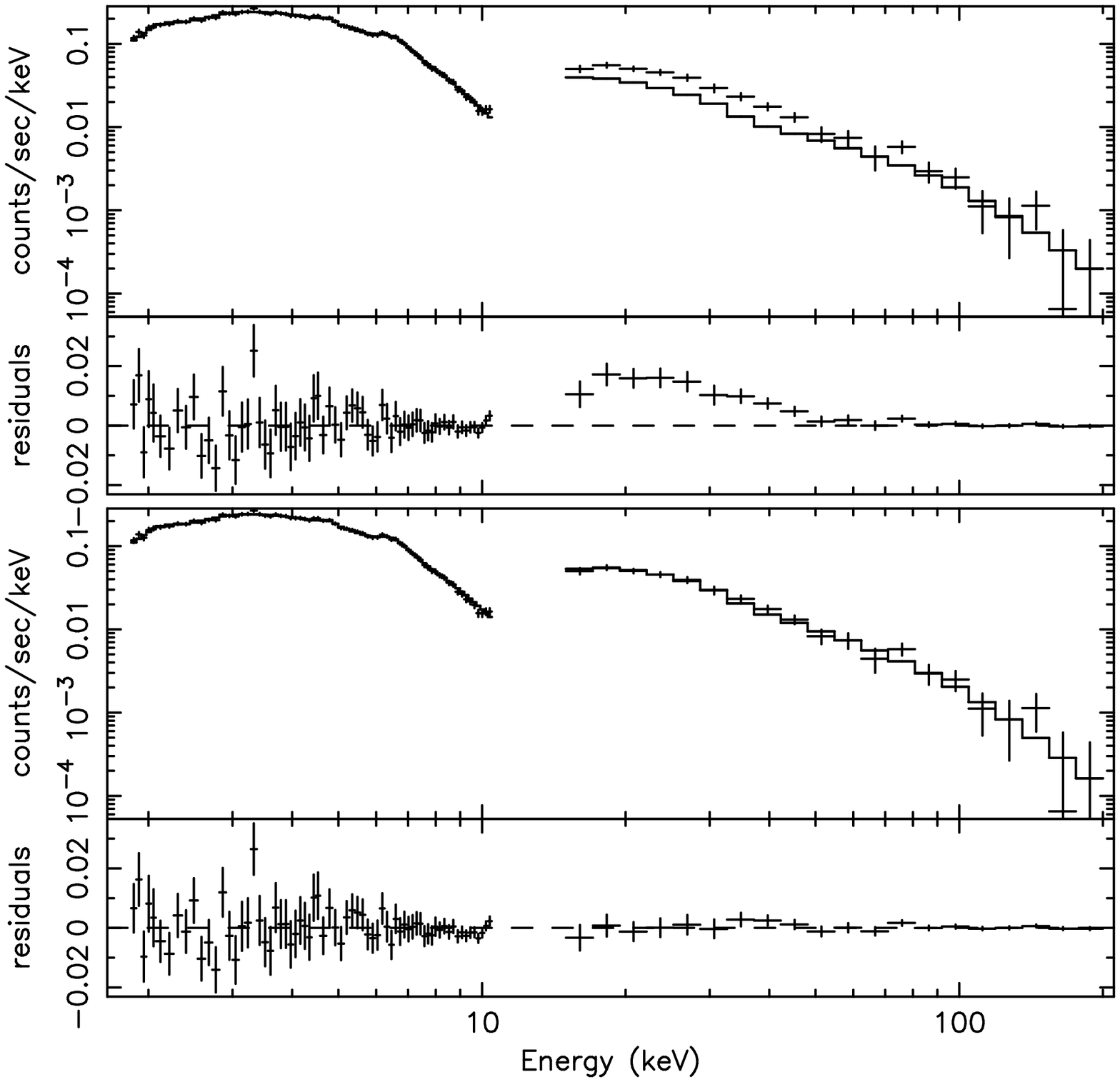,width=55mm,height=60mm,angle=-90}
\end{minipage}
\hspace*{0.5cm}
\begin{minipage}{58mm}
\epsfig{file=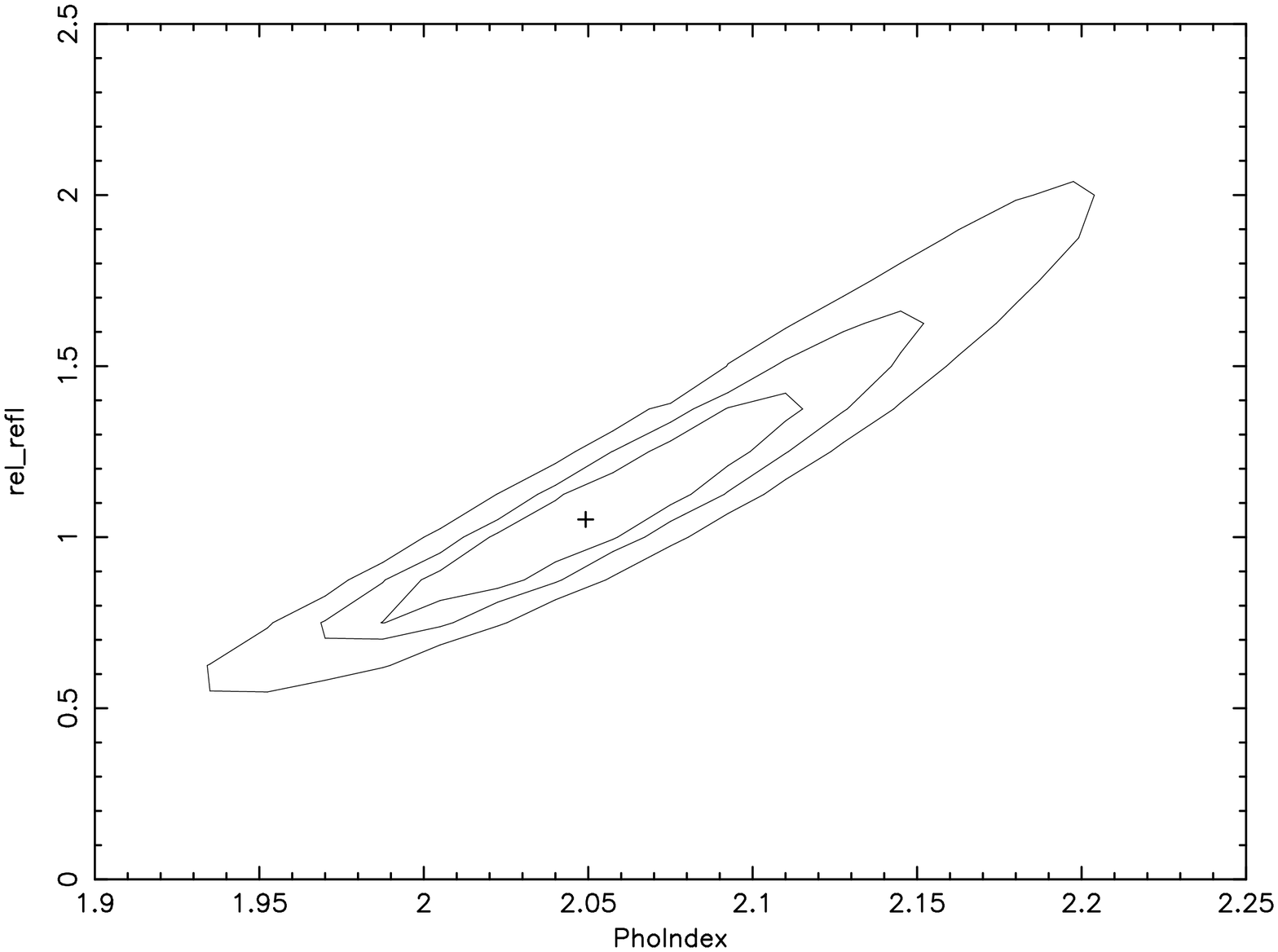,width=55mm,height=55mm,angle=-90}
\end{minipage}
\caption{Left panel: MECS and PDS data of NGC 5506 (Jan. 1997). In the upper
panel the MECS data alone are fitted with an absorbed power law plus gaussian
line and reflection component, and the best fit is extrapolated into
the PDS energy range. In the lower panel the same model is fitted to
MECS and PDS data together.
Right panel: for the same observation, confidence contours (68, 90, 99\%) of
the parameters $\Gamma$ and R.}
\label{5506a}
\end{figure}

\subsubsection{High energy cut--off.}

High energy cut--offs, expected in presently popular Thermal Comptonization
models and required to explain the hard X--ray background in terms of AGN
(e.g. Comastri et al. 1995), can also be searched for by
BeppoSAX if they are at low enough energies. For some of the sources in
Table 1 the results are rather inconclusive, due to the limited S/N. For
other sources more stringent lower limits are obtained to the $e$--folding
energy of an exponential cut--off.\footnote{Actual cut--offs in Thermal
Comptonization models may be sharper than an exponential law, as remarked to
me by A. Zdziarski. A conservative statement is that no deviations
from a power law up to $\sim$200 keV are required by the data.} For a few 
sources, finally, a positive detection is obtained. What it is clear from
the table is that there is no a ``universal" cut--off energy (i.e. temperature
in thermal models), but that the spread in this quantity is significant.

\subsubsection{Soft excesses.}

The low energy sensitivity of BeppoSAX down to 0.1 keV, coupled with the 
wide band which permits a reliable estimate of the intrinsic power law,
allows to search for soft excesses.\footnote{I mean ``true" soft excesses, 
to be distinguished from ``artificial" ones resulting from misfitting 
warm absorbers.} Surprisingly enough, no soft excesses have been detected
so far in classical Seyfert 1s (apart maybe from NGC~7469, 
this result being however preliminary: Piro, private communication). This
sharply constrasts with BeppoSAX results on Narrow Line Seyfert 1s, where
soft excesses are clearly observed, fully in agreement with both ROSAT and
ASCA findings (Comastri et al. 1998a,b; Brandt, this volume). 
Also in low redshift 
quasars soft excesses are usually observed by BeppoSAX (Fiore et al. 1998a).

While a quantitative comparison with previous 
results is still in progress, this promises to be one of the most intriguing
BeppoSAX result on Seyfert galaxies.

\section{Seyfert 2 galaxies}

According to unification models, 
Seyfert 2 galaxies are intrinsically identical 
to Seyfert 1s, but are observed through absorbing matter. 
Indeed, all Seyfert 2s detected so far
in X--rays show evidence for absorption in excess of the Galactic one. 
The appearance of a Seyfert 2 in X--rays depends strongly on
the column density of the absorbing matter. Two regimes may be identified:
Compton--thin and Compton--thick.
In the former case the absorbing matter has a column density lower
than 10$^{24}$ cm$^{-2}$, and is therefore optically thin to Compton 
scattering; the nuclear radiation is then directly visible above a few keV. 
In the latter case, i.e. when  $N_{\rm H}\simgt10^{24}$ cm$^{-2}$, the 
absorber is thick to Compton scattering: radiation is intercepted and, when
$N_{\rm H}\simgt10^{25}$ cm$^{-2}$, the nucleus is virtually
unobservable at all energies (up to the Klein--Nishina decline, at least),
because photons are downscattered to energies where 
photoelectric absorption dominates, and eventually get absorbed.
(For intermediate column densities, a
fraction of the nuclear radiation can still escape above $\sim$10 keV; 
the Circinus galaxy, described below, in one example). 
The obscuration of the nucleus, which in Compton--thick sources
is complete below 10 keV, permits to study in detail 
components which would have
otherwise been diluted by the nuclear emission. In particular,
reflection from both cold (the torus?) and ionized (the optical reflector?)
matter may become visible. Details of the continuum and line emission 
expected in Compton--thick Seyfert 2s may be found in e.g. Ghisellini, Haardt \&
Matt (1994), Krolik, Madau \& Zicky (1994), Matt, Brandt \& Fabian
(1996), Netzer (1996). 

Apart from a few scattered observations in the Guest Observer program, three
major Core Programs on Seyfert 2s are ongoing. The first one, led by 
L. Bassani, is a survey
of bright Compton--thin Seyfert 2s. A summary of the first results can 
be found in Bassani et al. (1998a). Among the five sources
observed, two of them (NGC~7674, Malaguti et al. 1998; Mrk~3,
Cappi et al. 1998) turned out to be Compton--thick. For the other, 
genuinely Compton--thin sources, broad band BeppoSAX observations
confirm the lack of any significant difference in X--ray properties
(apart of course from absorption) with Seyfert 1s, in agreement with
unification models.

The second and third programs, which I will discuss in more detail, 
are a survey of bright Compton--thick Seyfert 2s, and of an optically
selected sample of faint Seyfert 2s. 

\subsection{Bright Compton--thick Seyfert 2s}

Two objects have been observed so far in this program: 
Circinus Galaxy and  NGC~1068. 

\subsubsection{Circinus Galaxy.}

The Circinus galaxy is probably the nearest Active Nucleus, its 
distance being only $\sim$4~Mpc. It was observed by BeppoSAX on March 1998
(Matt et al. 1998b).
The spectrum below 10 keV is consistent with the ASCA one (Matt et al. 1996),
i.e. a reflection component from cold matter (including a huge iron K$\alpha$
line) and a soft excess of unclear origin. Other lines are also clearly
present in the spectrum (Guainazzi et al., in preparation), as evident from 
Fig.~3, again confirming ASCA findings. 

\begin{figure}
\centerline{\epsfig{file=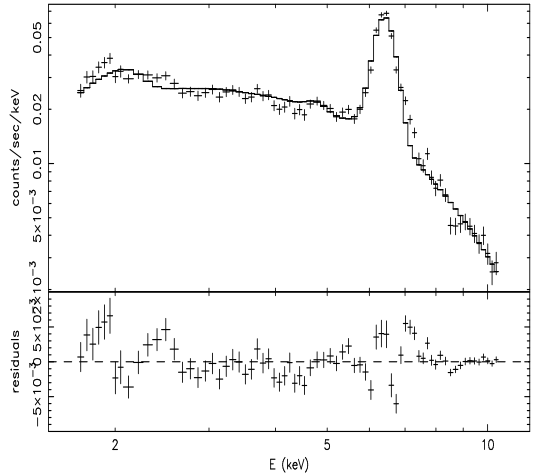,width=8cm,height=8cm,angle=-90}}
\caption{The MECS spectrum of the Circinus galaxy fitted with a power law
and an iron K$\alpha$ line. Note the other line features in the residuals.}
\label{circ_mecs}
\end{figure}

The surprise comes
when the best fit LECS+MECS model is extrapolated to the PDS band: it
dramatically falls short of the data (Fig.~4, left panel). No known
bright sources are present in the PDS field of view, so the best interpretation
of the spectrum in Fig.~4 is that the nuclear emission
is piercing through the absorber.
This implies that the column density of the absorber is greater than
$\sim$10$^{24}$ cm$^{-2}$ (to completely block the radiation below 10 keV)
but lower than $\sim$10$^{25}$ cm$^{-2}$ (to still permit 
significant transmission; see Ghisellini, Haardt \& Matt 1994). A similar
situation has already been observed in a couple of other sources, namely 
NGC~4945 (Iwasawa et al. 1993; Done, Madejski \& Smith 1996) and, with
BeppoSAX, Mrk~3 (Cappi et al. 1998). In this situation, pure absorber
models, like those usually available in {\sc xspec}, are no longer 
applicable, unless the covering factor of the absorbing matter is very
small. A model which includes Compton scattering is necessary; we have
then constructed one using MonteCarlo simulations 
and fitted the data with it. The resulting column density is about 
4$\times$10$^{24}$ cm$^{-2}$ and the 2--10 keV unabsorbed 
nuclear luminosity is $\sim 10^{42}$ erg s$^{-1}$. 

\begin{figure}
\begin{minipage}{60mm}
\epsfig{file=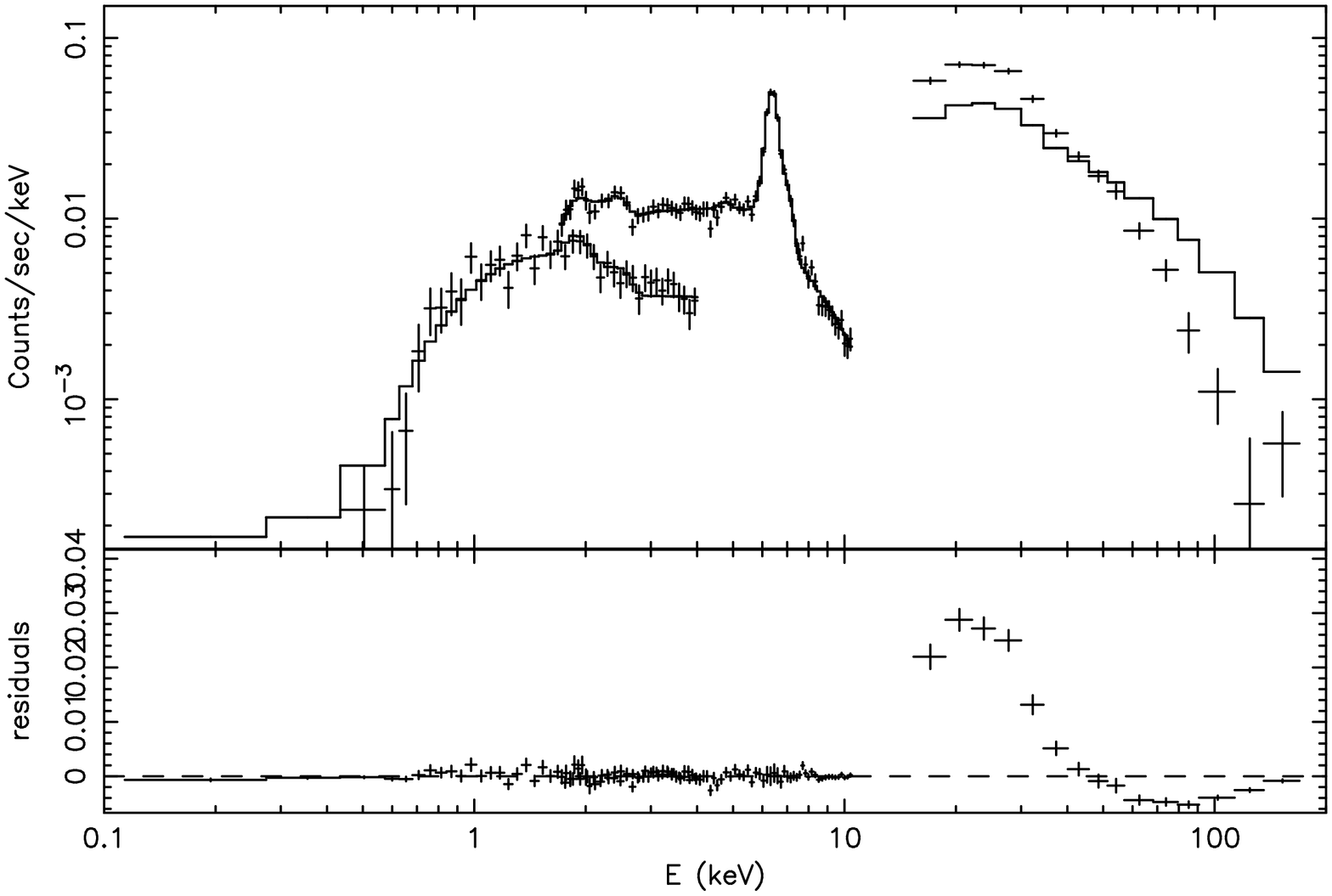,width=55mm,height=60mm,angle=-90}
\end{minipage}
\hspace*{0.5cm}
\begin{minipage}{60mm}
\epsfig{file=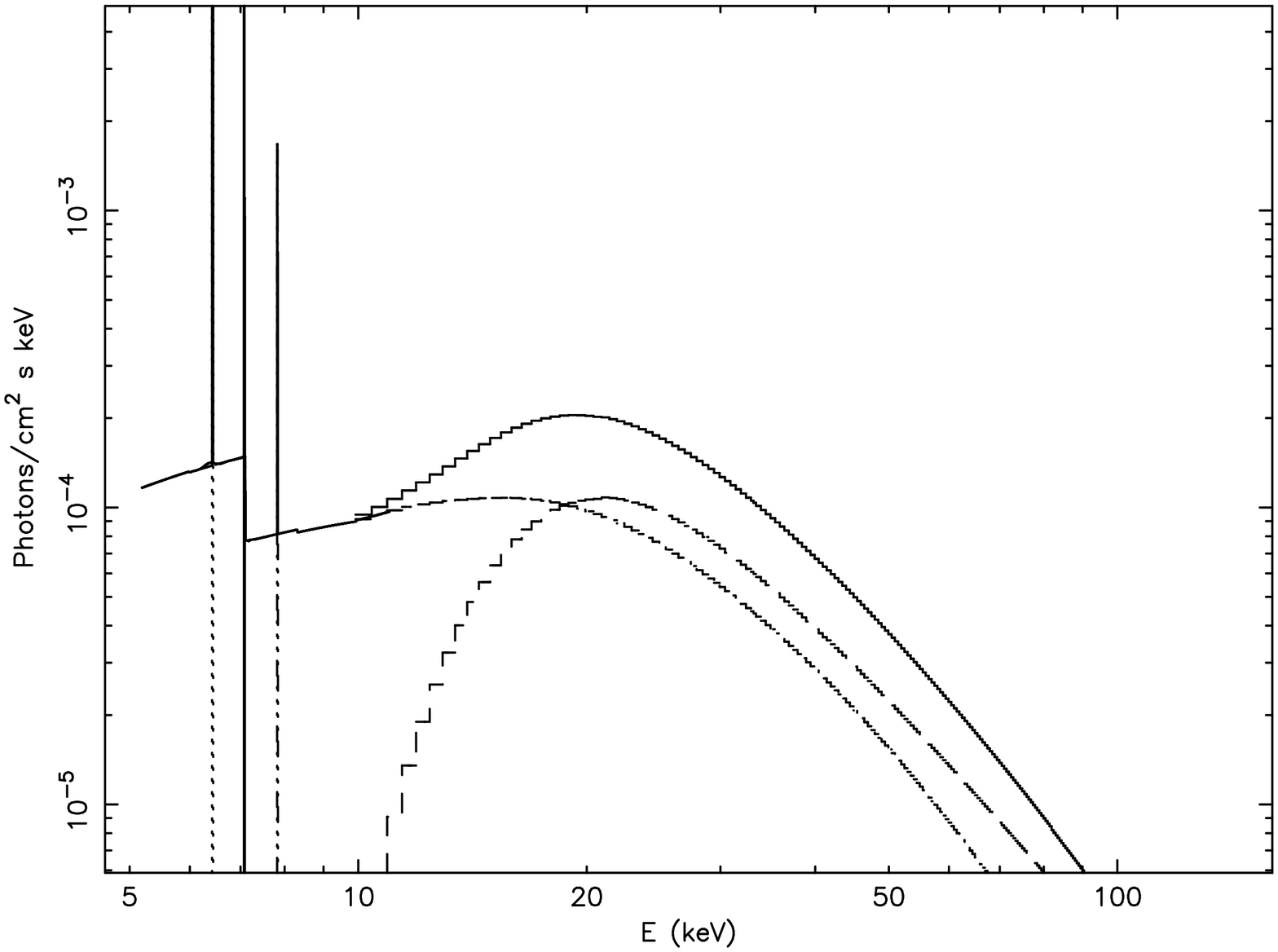,width=55mm,height=60mm,angle=-90}
\end{minipage}
\caption{Left panel: the LECS+MECS spectrum of the Circinus galaxy extrapolated
to the PDS band. Excess emission above 10 keV is evident. Right panel:
best fit model composed by a reflection and a transmission component, plus
K$\alpha$ and K$\beta$ iron lines, and K$\alpha$ nickel line.}
\label{circ_spec}
\end{figure}

\subsubsection{NGC 1068.}

It was observed around New Year 1997 for about 100 ksec (effective
MECS observing time), and then again one year
later for about 37 ksec. Results from the first observation
on the high energy part of the spectrum can be found in Matt et al. (1997).
The full band analysis of both observations is reported in Guainazzi et al.,
in preparation. 

The broad band spectrum is rather complex: above about 4 keV the spectrum
is dominated by reflection of the nuclear radiation from both
cold matter, which  dominates above $\sim$10 keV, and from warm matter. 
The presence of two reflection components was already noted by Marshall
et al. (1993) in the BBXRT data, which in fact revealed
a complex iron line with both neutral and highly ionized  lines,
a results later confirmed and improved by ASCA observations 
(Ueno et al. 1994). Matt, Brandt \& Fabian (1996) and 
Iwasawa, Fabian \& Matt (1997) attributed the cold reflection to
the inner surface of the torus, and the highly ionized (He--like and 
H--like) lines to fluorescence and resonant scattering from optically
thin (but possibly thick to resonant absorption), photoionized material.
BeppoSAX, thanks to the hard X--ray sensitivity provided by
the PDS, has permitted to confirm this scenario and estimate the relative
contributions of the two components (Matt et al. 1997; see 
Fig.~5). 

\begin{figure}
\begin{minipage}{58mm}
\epsfig{file=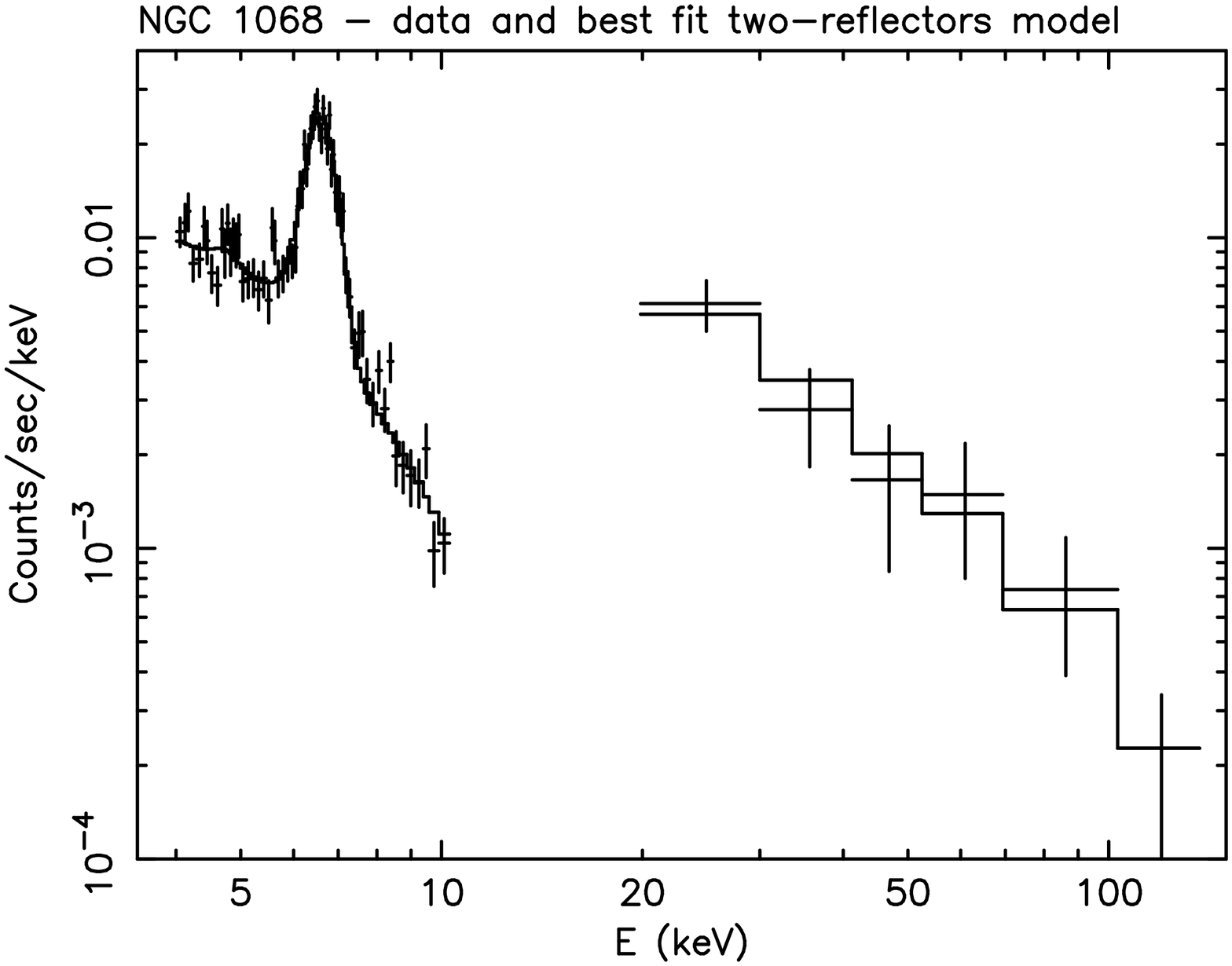,width=55mm,height=58mm,angle=-90}
\end{minipage}
\hspace*{0.5cm}
\begin{minipage}{58mm}
\epsfig{file=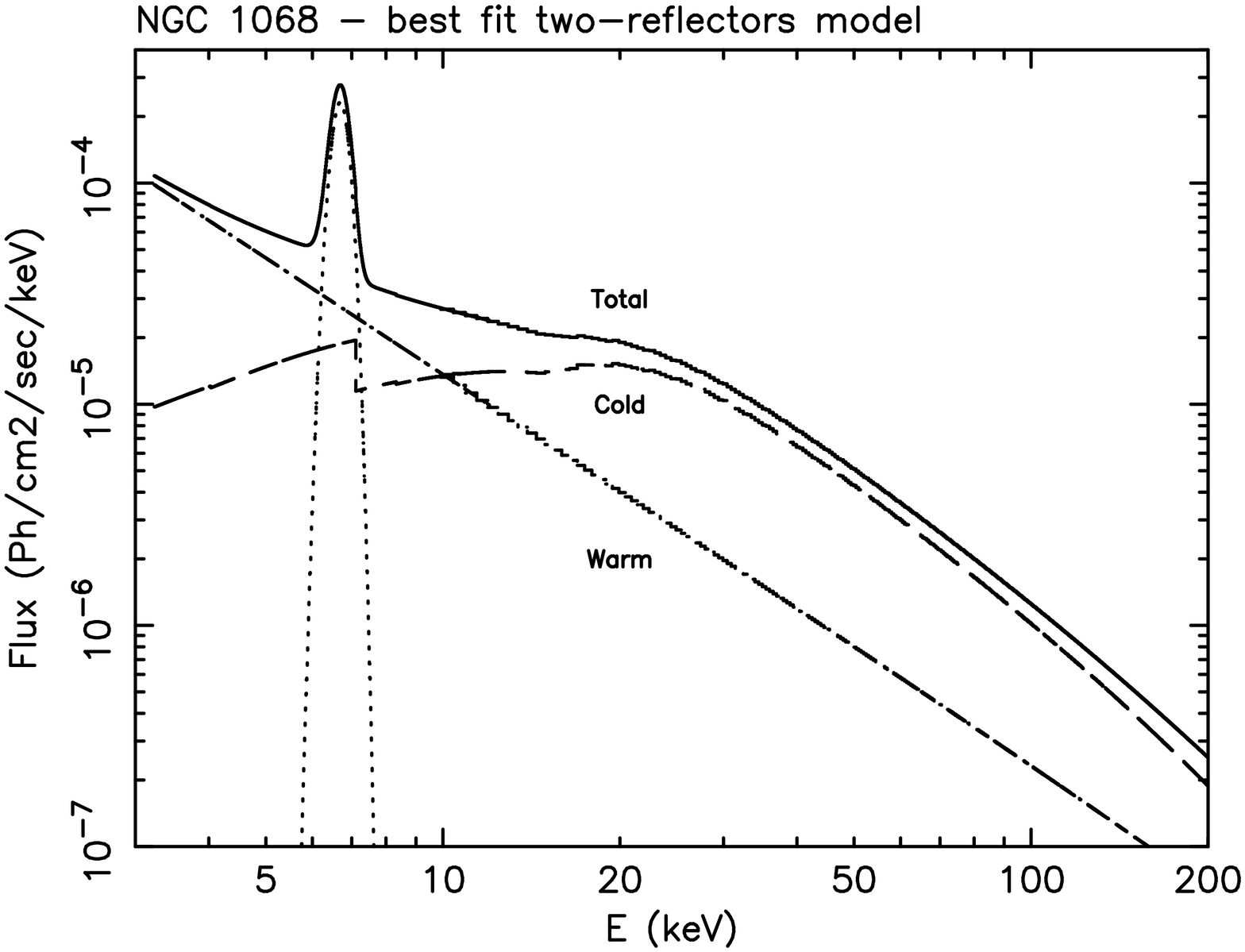,width=55mm,height=55mm,angle=-90}
\end{minipage}
\label{hard1068}
\caption{Spectrum (left panel) and best fit model (right panel) of NGC~1068.
The model is composed by a broad iron line (actually a blend of different
lines) and two reflectors, one neutral and optically thick, the other
highly ionized and optically thin. From Matt et al. (1997).}
\end{figure}

Below 4 keV, the spectrum is well described by a thermal--like component,
probably related with the starburst regions (Wilson et al. 1992), plus
the warm reflection component. As already known from ASCA
(Ueno et al. 1994; Turner \& Netzer 1997), the 
spectrum is very rich in emission lines (Fig.~6).
A detailed discussion on the line spectrum 
can be found in Guainazzi et al., in preparation.

\begin{figure}[t]
\centerline
{\epsfig{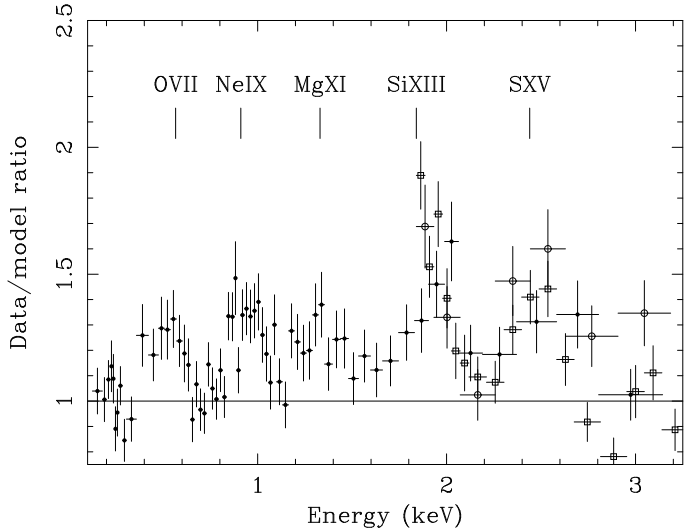}}
\label{1068_lines}
\caption{Data/model (i.e. warm reflector + thermal plasma) for NGC~1068.
Many lines are apparent. The most likely identification
for each line is indicated. From Guainazzi et al., in preparation.}
\end{figure}

Long term variability has also been searched for. 
The 0.1-1 keV LECS light curves are shown in Fig.~7, together
with the background.
The left panel refers to the first observation, the right panel
to the second one.  A variation between the two observations is
apparent, which clearly cannot be
ascribed to variations in the background. The 
gain is also very stable. Unfortunately, however, it is 
possible that such  variations arise from a larger 
shadowing, in the second observation, from the grid on top of the detector: 
even if the nominal position of the source in the two observations 
does not allow for such a change, uncertanties in the
aspect reconstruction cannot permit to rule out definitely this hypothesis.
The variation, if true, would be very important, 
as it would permit to estimate the size of the emitting
region to be at most 1 light year. The starburst component is very 
extended  (Wilson et al. 1992), and therefore 
the variation should be ascribed to the
warm reflection component. Further long term monitoring of the source
will hopefully settle this issue. 

\begin{figure}[t]
\centerline
{\epsfig{file=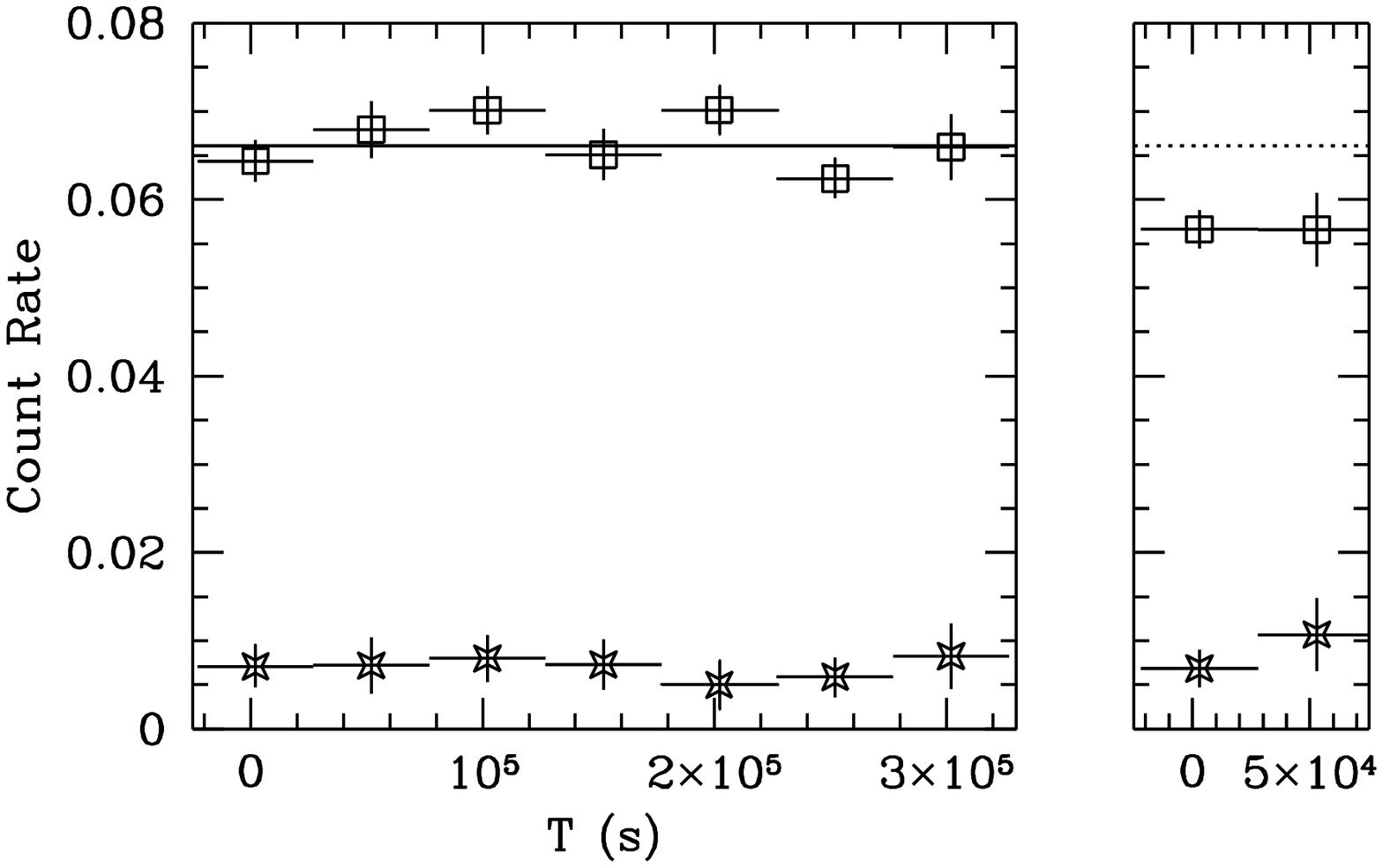,width=9cm,height=7cm,angle=0}}
\caption{0.1--1 keV LECS light curve of NGC~1068. Both source and
background light curves are shown.
Left (right) panel refers
to the first (second) observation (see text). From 
Guainazzi et al., in preparation.}
\label{1068_var}
\end{figure}

\subsection{An optically selected sample of Seyfert 2s, and the
unification model}

The results described above can be easily understood in the framework
of unification models. However, in recent years many works 
showed that some intrinsic differences bewteen the average properties of
Seyfert 1s and 2s do exist: enhanced star formation
in Seyfert 2 galaxies (Maiolino et al. 1997); different
morphologies between galaxies hosting type 1 and 2 nuclei, those hosting
type 2 being on average more irregular (Maiolino et al. 1997, Malkan et al.
1998); a greater dust content in Seyfert 2s
(Malkan et al. 1998).
The aspect angle is clearly not the only relevant parameter. The next
question is, of course, whether the differences involve 
the nuclear properties or only the environment.  In this respect, it
is important to
observe in hard X--rays (where nuclear activity dominates the emission)
a sample of optically selected sources. Salvati et al. (1997) and Maiolino 
et al. (1998) studied with BeppoSAX
an OIII flux--limited sample of Seyfert 2s,
in the assumption that the OIII flux is a good isotropic 
indicator of Seyfert activity (this is probably not completely true, 
but it is in any case the best one). 
The first result of this program is that all sources observed
so far have been detected (9 out of 9, Risaliti et al. 1998), 
with typical X--ray luminosities exceeding those of normal galaxies. This is 
a rather strong indication that {\sl all Seyfert 2 have a type 1
nucleus}, and then that any difference between the two classes should
be searched for in the nuclear environment. Let me suggest a modification
to the zero--order unification model
(see Fig.~8): all Seyferts have a type 1 nucleus plus
circumnuclear, $\sim$100 pc--scale
dust lanes (Malkan et al. 1998), which I presume to be
optically thin to Compton scattering (otherwise, too much circumnuclear
matter would probably be present).
Only a fraction of Seyferts, however, have also the (Compton--thick)
molecular torus, which possibly forms preferentially
in irregular, disturbed galaxies
(which have also, probably for the same reason, an enhanced star formation
activity as well as an overall greater dust content). As illustrated
in  Fig.~8, if the nucleus 
is directly observed, the source is a Seyfert 1. If the line--of--sight
intercepts matter other than the torus (i.e. a dust lane, or even the
galactic disc for highly inclined galaxies) the source falls in the
mixed bag including Compton--thin
Seyfert 2's, intermediate Seyfert and NELGs. 
If, finally, the line--of--sight
intercepts the torus, the source is a Compton--thick Seyfert 2 galaxy. 
In this scenario, Seyfert 1s would be preferentially, but not exclusively
(see next section for a Seyfert 1 {\sl with} the torus) observed among 
torus--free sources. 
It is worth noticing that another results of the program 
under discussion is that
Compton--thick sources contributes to a large fraction of the total 
(see Fig.~9).
Moreover, other Compton--thick sources have been discovered by BeppoSAX
(Malaguti et al., 1998; Ueno et al. 1998; Cappi et al. 1998).
Therefore, even if in the
proposed scenario the presence of a molecular torus is no longer
ubiquitous, it should nevertheless still be rather common (see also below). 

\begin{figure}[t]
\centerline{\epsfig{file=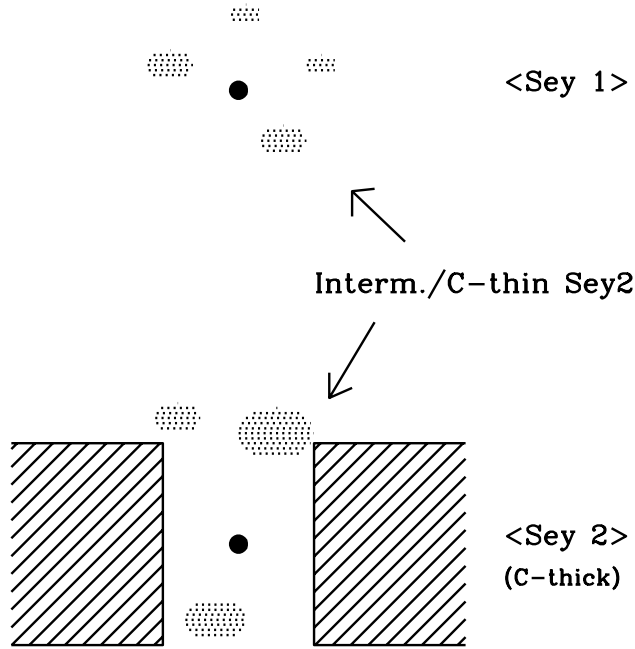,width=11cm,height=11cm}}
\caption{My best guess for the unification model (see text): the typical
Seyfert 1 does not possess a torus (but a fraction Seyfert 1s with 
a torus are also expected), which is a prerogative 
of the typical Compton--thick Seyfert 2s. Intermediate Seyfert 
and Compton--thin
Seyfert 2s are those objects observed through optically thin, to Compton
scattering, dust lanes}. 
\label{matt_um}
\end{figure}

One of the aim of the program is to determine the distribution of column
densities. As shown in Fig.~9, 
a large fraction of sources turned out to be Compton--thick. However, the
sample is biased in favour of high column densities. In fact, sources
have been selected on the basis of the OIII flux, but excluding sources
for which good spectra (from Ginga and/or ASCA) were already available. 
The Ginga and ASCA samples were largely X--ray selected, and therefore
biased towards low column density
(and therefore large flux). The real N$_{\rm H}$ distribution should then
be a mixture of that from BeppoSAX and the ones from Ginga and ASCA. 
Collecting all data available in the literature, Bassani et al. (1998b)
have derived the ``true" column density distribution: the average
value of the column is $\sim$3$\times10^{23}$ cm$^{-2}$, and the 
fraction of Compton-thick sources is as large as 30\%. 

\begin{figure}
\begin{minipage}{65mm}
\epsfig{file=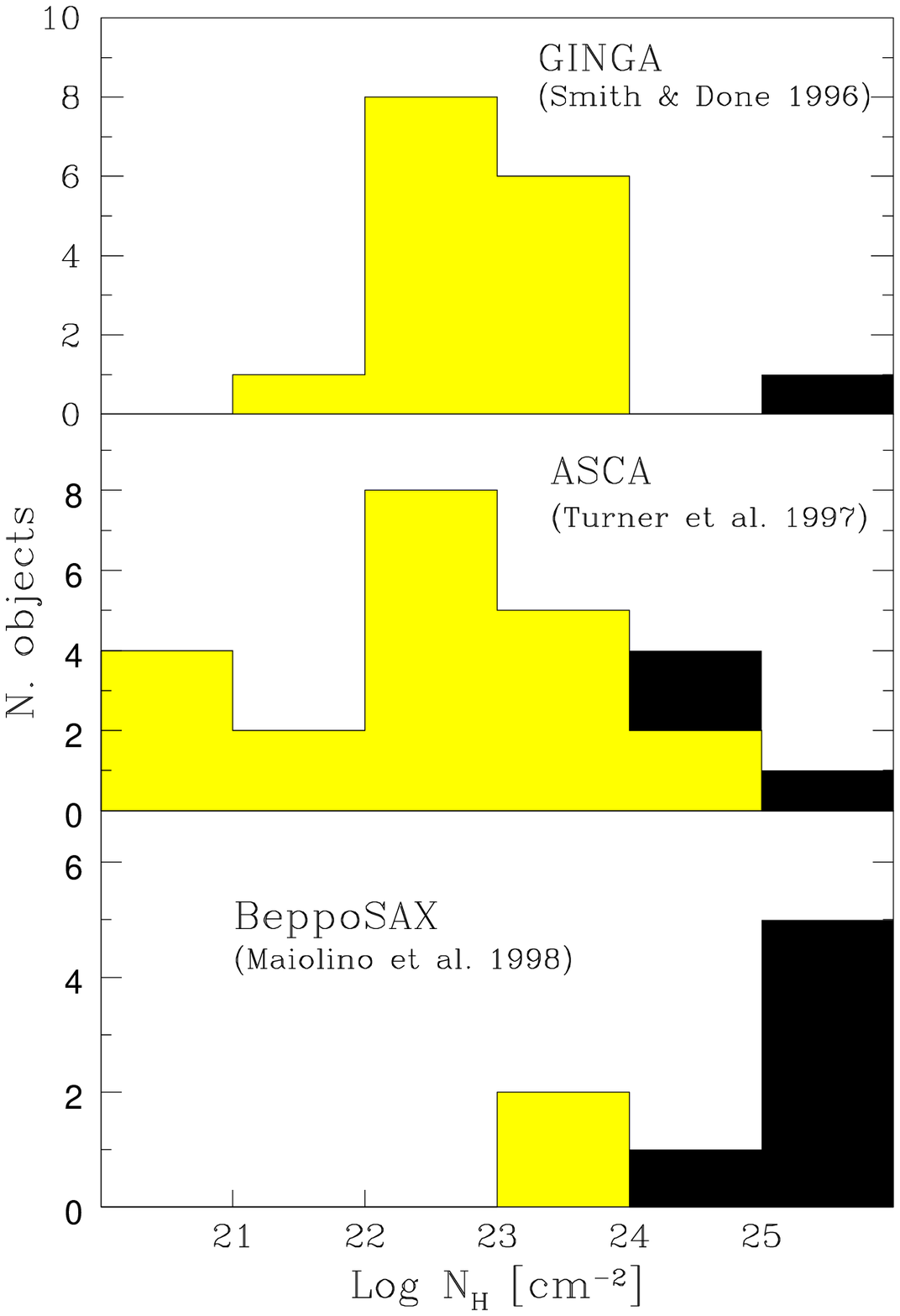,width=65mm,height=70mm,angle=0}
\label{matt_maio1}
\caption{ BeppoSAX distribution of column densities, compared with the 
Ginga and ASCA results. From Maiolino et al. (1998).}
\end{minipage}
\hspace*{0.cm}
\begin{minipage}{65mm}
\epsfig{file=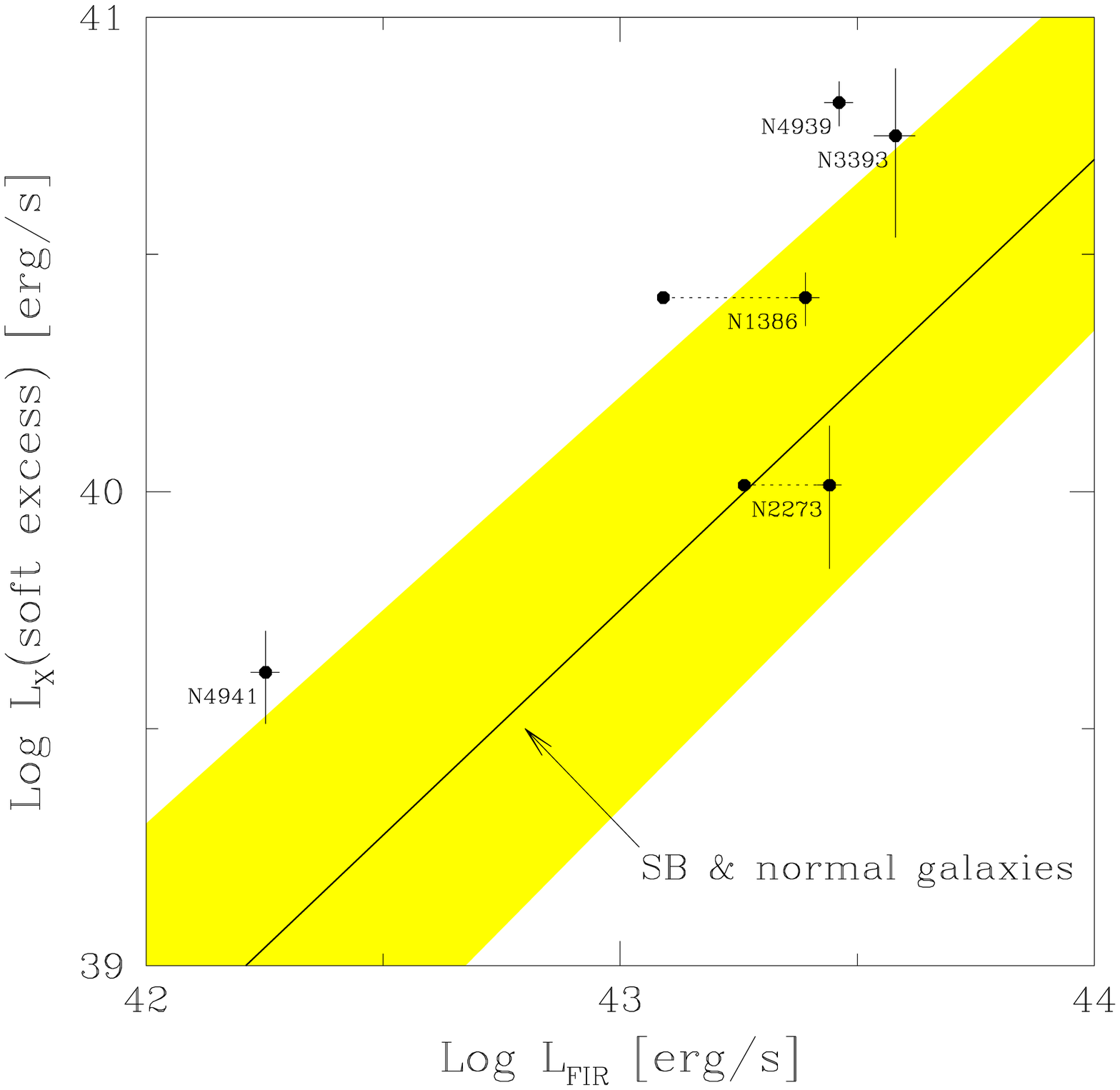,width=65mm,height=55mm,angle=0}
\caption{ Distribution of the 0.5--4.5 keV excess emission vs. the FIR
luminosity for the sources in the Maiolino et al. (1998) sample. The shaded
area indicates the relation for starburst and normal galaxies.}
\label{matt_maio2}
\end{minipage}
\end{figure}

Most of the sources in the Maiolino et al. (1998) sample show evidence
for soft X--ray emission in excess of the (absorbed) hard X--ray emission.
The question is whether this emission is related to nuclear activity
(i.e. scattering of the nuclear radiation from ionized material) or is
rather related to starburst regions. In Fig.~10 the soft X--ray
luminosities vs. the IR luminosities of the sources in the sample, for which 
a good estimate of the soft excess is possible, are shown. The thick solid line 
indicates the relationship for starburst and normal galaxies 
(David et al. 1992); the shaded
area is the 90\% confidence limit on this relation. Clearly, more data are
needed, but at a first glance one would say that in general 
some nuclear--related
emission contributes, but not necessarily dominates, the soft X--rays.

\section{The strange case of NGC~4051}

\begin{figure}
\begin{minipage}{60mm}
\epsfig{file=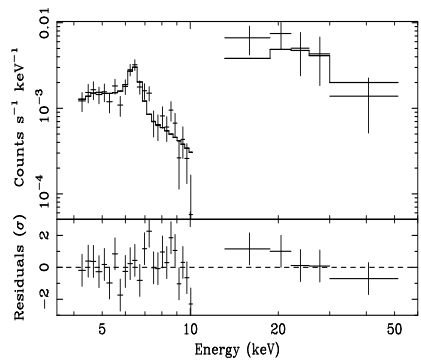,width=55mm,height=60mm,angle=-90}
\end{minipage}
\hspace*{0.5cm}
\begin{minipage}{60mm}
\epsfig{file=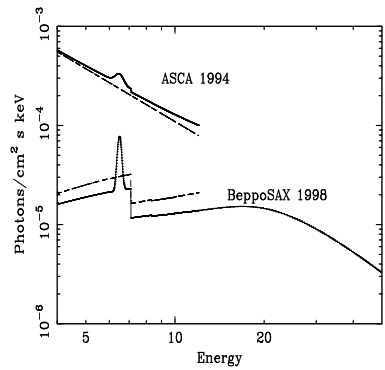,width=55mm,height=60mm,angle=-90}
\end{minipage}
\caption{Left panel: May 1998 BeppoSAX spectrum of NGC~4051. Right panel:
bets fit model, compared with the 1994 ASCA one. From 
Guainazzi et al. (1998c)}
\label{4051}
\end{figure}

The spectrum in the left panel of Fig.~11
(from Guainazzi et al. 1998c) closely resembles those
of Compton--thick Seyfert 2 galaxies (see previous section), i.e. a very flat
continuum and a prominent iron line. And, actually, the interpretation is
the same: a reflection--dominated spectrum. The surprising fact is that
this  is the spectrum of a well known Seyfert 1 galaxy,   
NGC~4051. The source was observed by BeppoSAX on 9-11 May 1998, and was
caught in an unprecedented low state (see right panel of  Fig.~11), 
which show the BeppoSAX best fit model together with 
the 1994 ASCA one), 20 times fainter than the average
flux. The source is usually very variable on fairly short time scales, and 
in the past it has probably been observed at such a low flux level, but
usually only for a few thousands seconds, never for a so long time. 
The upper limit on the
nuclear component implies a dimming of at least a factor 35 with respect
to the average value.
We have therefore observed, for the first time, what can be practically 
considered a switching--off of an active nucleus, which
has left the reflection component as the only echo of the past activity.
This may also be considered one of the strongest evidence for the
presence of substantial circumnuclear matter in Seyfert 1s.

Comparing the BeppoSAX and ASCA best fit models in the right panel of 
Fig.~11, it seems that also the reflection component was lower
during the BeppoSAX observation, even if this evidence must be taken with
great caution. This could imply that part of the reflection, when the nucleus
is active, originates in the accretion disc, which would respond almost
simultaneously (say, within $\sim$1 day) to any variation of the nuclear
radiation. Alternatively, if all 
the reflection comes from a pc--scale torus (or sub--pc; remember
that NGC~4051 is one of the less luminous AGN, and it is well possible
that all distances are scaled down there),
a fading of the reflection component on time scales of months is expected.
This would have provided a tool to ``measure" the size of the torus.
We therefore observed the source again about one and half month later.
Unfortunately, this time the source did not collaborate: 
it resumed the normal level of activity, and the spectrum was no longer
reflection--dominated (Orr et al. 1998).

\section{The HELLAS sample and the XRB}

The good spatial resolution of the 
MECS instrument onboard BeppoSAX has been exploited to 
search for serendipitous, hard X--ray sources in the field of view
of pointed targets (Giommi et al. 1998).
One of the aims of this program, called HELLAS (High Energy Llarge Area 
Survey), is to assess, in a band 
where its energy density is much greater than 
in the so far most studied band below $\sim$2 keV, the nature
of the sources making the X--ray Background (XRB;
see Barcons 1996 and Matt 1995 for recent reviews).
The survey presently consists of
about 150 sources in the 5--10 keV energy band (a fraction of
them being detected {\it only} in this band) over an area of about 50
degrees. Taking properly into account the sky coverage, these numbers
translate to a density of $\sim$ 20 sources per square degree down to
a 5--10 keV flux of 5$\times10^{-14}$ erg cm$^{-2}$ s$^{-1}$, which in
turn corresponds to a 35-40 percent of the cosmic
X-ray background in this band being resolved (Fiore et al. 1998b). 

The next question, of course, is the nature of these sources. An extensive
identification program is in progress, with several night allocated
to both northern and southern telescopes. First runs 
have already been highly rewarding, as reported in Fiore et al. (1998b).
Out of ten sources observed (for all of them specta were taken of 
all source, within the MECS error box, down to a R magnitude
of 20), 9 of them have been identified. Note that the sources were selected
as being the only ones visible from  the telescope at that time of the year,
and therefore there is no astrophysical bias whatsoever. 
8 sources were identified with AGNs. (The eight source is identified with a 
LINER, therefore a borderline object). 
The corresponding chance probability is only 0.15\%. It is worth noticing
that the averege spectrum of these sources is consistent with that of the
XRB. Therefore, from all the pieces of evidence just summarized, one is
authorize to infer that at least a large fraction 
of the XRB in this band is made by AGN. 

Of the 8 AGN identified, 3 are normal, ``blue" QSO, 2 are ``red" QSO,
and 3 are intermediate (1.8--1.9) Seyferts, with redshifts ranging from
0.17 to 1.3. Four sources (the three intermediate Seyferts and one
of the red quasars) show evidence for X--ray absorption with column 
densities in  excess of 10$^{22}$ cm$^{-2}$. These findings therefore 
support models 
in which the hard XRB is made by a mixture of obscured and unobscured 
AGN (Comastri et al. 1995, and references therein).

\acknowledgments
I thank all my collaborators in the various BeppoSAX observing programs 
described in this paper. In particular, 
I thank F. Fiore, M. Guainazzi, R. Maiolino, F. Nicastro, 
and G.C. Perola for many helpful
discussions, and L. Piro for allowing me to quote results before 
publication. I am indebted to the BeppoSAX Science Data Center for frequent
help in the data reduction. I acknowledge financial support
from ASI and MURST.

\end{document}